%% file: 1550_lags.tex
\documentstyle[12pt,aasms4]{article}

\begin{document}

\title{Characterizing the Low-Frequency QPO behavior of the X-ray Nova XTE~J1550--564}
\author{Gregory J. Sobczak\altaffilmark{1}, 
Ronald A. Remillard\altaffilmark{2}, Michael P. Muno\altaffilmark{2}, 
Jeffrey E. McClintock\altaffilmark{3}}

\altaffiltext{1}{Harvard University, Astronomy Dept., 60 Garden St. MS-10, Cambridge, MA
02138; gsobczak@cfa.harvard.edu}
\altaffiltext{2}{Center for Space Research, MIT, Cambridge, MA 02139; rr@space.mit.edu,
muno@space.mit.edu}
\altaffiltext{3}{Harvard-Smithsonian Center for Astrophysics, 60 Garden St. MS-3, Cambridge, MA
02138; jem@cfa.harvard.edu}

\begin{abstract} 

For all 209 RXTE observations of the 7-Crab X-ray nova XTE~J1550--564, we analyzed the
X-ray power spectra, phase lags, and coherence functions.  These observations cover
the entire 250 day outburst cycle of the source and are the most complete and arguably
the richest data set for any black hole X-ray nova.  We find that there are three
fundamental types of QPO behavior--one more than reported by Wijnands, Homan, \& van
der Klis (1999, \apj, 526, 33).  The new type occurred during the first half of the
outburst.  These three types of QPO behavior can be grouped according to the relative
contributions of the disk and power-law components to the total flux.  

\end{abstract}


\section{Introduction}

XTE~J1550--564 is an X-ray nova and black hole candidate discovered with the All Sky
Monitor (ASM; Levine et al.~1996) onboard the {\it Rossi X-ray Timing Explorer} (RXTE)
in 1998 September (Smith et al.~1998).  The outburst lasted until 1999 May; its
duration was approximately 250 days, during which time RXTE performed an almost daily
series of pointed observations.  A summary of these observations can be found in
Sobczak et al.~(2000b).  Extensive spectral and timing studies of this source have
been performed using these data.  The source has been observed in the very high,
intermediate, and high/soft outburst states of black hole X-ray novae, with an X-ray
spectrum that can be well modeled by multicolor blackbody disk and power-law
components (Sobczak et al.~1999,2000b; Homan et al.~2000).  

The power spectra of XTE~J1550--564 exhibit quasiperiodic X-ray oscillations (QPOs)
at both high frequencies ($\sim100$--285~Hz) and low frequencies (0.08--18~Hz)
(Remillard et al.~1999; Sobczak et al.~1999; Cui et al.~1999; Homan et al.~2000). 
Herein we are interested only in the latter `low-frequency' QPOs.  These QPOs can
have large amplitudes, with peak to trough ratios as high as 1.5 (Remillard et
al.~1999), which indicates that they are a fundamental aspect of the accretion
process.  A sample lightcurve of XTE~J1550--564 can be seen in Figure~1 of Sobczak et
al.~(2000a), which explored the correlations between the QPO frequency and amplitude
and the X-ray spectral parameters.  They reported that both the power-law and the
disk components are linked to the QPO phenomenon: QPOs are observed only when the
power-law component contributes more that 20\% of the 2--20~keV flux, and the QPO
frequency generally increases as the disk flux increases.

Our focus in this paper is on the properties of the phase lags associated with the
low-frequency QPOs.  The Fourier time delay, or lag, can be computed to measure the
phase lags between soft and hard X-ray energy bands.  Recent studies have shown that
the phase lags associated with low-frequency QPOs and their harmonics are quite
complex.  In one observation of GRS~1915$+$105, Cui (1999) found that the lower energy
(soft) photons lagged behind the higher energy (hard) photons (``soft lag'') for the
fundamental QPO and it's 2nd harmonic; however, the higher energy photons lagged
behind the lower energy photons (``hard lag'') for the 1st and 3rd harmonics. 
Wijnands, Homan, \& van der Klis (1999) reported even more complicated behavior from
14 observations of XTE~J1550--564 near the end of the outburst.  However, Wijnands et
al.~were able to show that the complicated phase lag behavior observed in these 14
observations actually consists of variations on only two fundamental types.  Phase
lags for 13 observations during the initial rise of XTE~J1550--564 are discussed by
Cui, Zhang, \& Chen~(2000).  

In this paper we analyze the phase lags for all 209 RXTE observations of
XTE~J1550--564.  These observations cover the entire 250~day outburst cycle of the
source and are the most complete and arguably the richest data set for any black hole
X-ray nova.  We find that there are three fundamental types of phase lag behavior --
one more than reported by Wijnands et al.~(1999).  The new type occurred during the
first half of the outburst.  We discuss correlations between the phase lags and the
X-ray spectral parameters, QPO frequencies, and amplitudes.

\section{Observations and Data Analysis}

The PCA data were accumulated with $125\mu$s resolution in `single bit' mode,
providing two energy channels corresponding roughly to 2--6 and 5--13~keV (modes
SB\_125us\_0\_17\_1s \& SB\_125us\_18\_35\_1s), respectively.  Above 13~keV, the event
mode was used with 16 energy channels and $16\mu$s time resolution (mode
E\_16us\_16B\_36\_1s).  The single bit data modes were combined into a single channel,
resulting effectively in two energy channels covering 2--13~keV and 13--30~keV (which
corresponds to PCA channels 0--35 and 36--255, respectively).  These three data modes
and channels ranges were not available for five of the observations listed in Table~1.
The alternate channel ranges for those observations are given in the footnotes to
Table~1.

We compute the phase lag between the two energy bands as follows (Bendat \& Piersol
1986; Cui et al.~1997; Vaughan \& Nowak 1997).  Let $a(t)$ \& $b(t)$ denote the
lightcurves in two energy bands with Fourier transforms $A(f)$ \& $B(f)$.  The cross
spectral density or cross power spectrum is given by
\begin{equation}
C(f) = \langle A^*(f) B(f) \rangle,
\end{equation}
where $\langle \rangle$ denotes an average over on ensemble of measurements.  If
$R(f)$ and $I(f)$ are the real and imaginary parts of $C(f)$, then the average phase
lag of $b(t)$ with respect to $a(t)$ is
\begin{equation}
\Delta \phi(f) = \tan^{-1} \left( \frac{I(f)}{R(f)} \right),
\end{equation}
It follows that the time lag of $b(t)$ with respect to $a(t)$ is $\Delta t = \Delta
\phi/2 \pi f$.  

The coherence function, $\gamma^2(f)$, is a measure of the linear correlation 
between two simultaneous time series as a function of frequency and is defined by 
\begin{equation}
\gamma^2(f) = \frac{ \vert \langle C(f) \rangle \vert^2 }{\langle \vert A(f) \vert^2 \rangle
\langle \vert B(f) \vert^2 \rangle}
\end{equation}
(Bendat \& Piersol 1986; Vaughan \& Nowak 1997; Nowak et al.~1999).  The coherence
functions reported here include a correction for the effects of Poisson noise and were
calculated using equation~(8) in Vaughan \& Nowak (1997).  If the coherence is
significantly less than unity then the measured phase lag cannot be trusted.  

See Remillard et al.~(1999) and Sobczak et al.~(2000a) for information regarding the
analysis of the power density spectra (PDS) for XTE~J1550--564.  The QPO properties
for each observation are listed here in Table~1.

\section{Results}

We find three fundamental types of phase lag behavior for XTE~J1550--564 -- one more
than the two types identified by Wijnands et al.~(1999).  Figures~1--4 show
representative power density spectra, phase lags, and coherence functions for these
three types of timing behavior.  Types A and B correspond to those identified
previously by Wijnands et al.  Type C, the third type of phase lag behavior, appears
during the first half of the outburst that was not analyzed by Wijnands et al. 
Figure~6 shows that these three types of QPO behavior can be grouped according to the
relative contribution of the disk (or power-law) component to the total flux.  The
characteristics of each type of QPO behavior are discussed below and summarized in
Table~2.  

Type A timing behavior is characterized by broad QPO features in the PDS ($Q \sim
2$--3, where $Q=\nu/\Delta \nu$) near 6~Hz that are likely the superposition of a QPO
peak and it's harmonic (Fig.~1).  The integrated rms amplitude of the QPO features is
a few percent.  The phase lags for type A behavior are generally featureless except
for a broad soft lag centered near the QPO feature with $\vert \Delta \phi \vert \sim
1$~rad and poor coherence ($< 50$\%).  The soft spectral component from the accretion
disk contributes $\sim$~50--70\% of the 2--20~keV flux during type A behavior
(Fig.~6).  We identify 4~observations with type A QPO behavior, all of which display
high-frequency ($\nu \sim$~100--284~Hz) QPOs (Remillard et al.~1999; Homan et
al.~2000; Remillard, private communication).  

Type B timing behavior is characterized by a narrow fundamental QPO feature ($Q \sim
10$) at 5--6~Hz (Fig.~2) with rms amplitude $\sim4$\%.  The first and sub-harmonic of
the fundamental QPO feature are also apparent at 11--12 and 2.5--3~Hz,
respectively.  Type B observations display a hard lag nearly coincident with, but
slightly off-center from, the fundamental QPO feature with $\Delta \phi
\sim 0.0$--0.4~rad, as well as soft lags associated with the first and sub-harmonics. 
The coherence of the type B fundamental lags is typically within 10\% of unity and has
small statistical errors, but the coherences of the harmonics have large errors.  Type
B timing behavior is observed when the power-law component contributes 60--75\% of the
2--20~keV flux (Figure~6).  We identify 9~observations with type B QPO behavior, all
of which display high-frequency QPOs.

We now introduce type C timing behavior (Fig.~3), which occurs primarily during the
first half of the outburst when the source is in the very high state and the power-law
component contributes $\gtrsim 75$\% of the 2--20~keV flux (Figure~6).  Type C timing
behavior is characterized by sharp ($Q \gtrsim 10$) fundamental QPO features with a
range of rms amplitudes from 3--16\%.  The first harmonic is present, and the first
sub-harmonic is usually observed as well.  The phase lags for type C behavior are
typically modest, with the fundamental exhibiting soft lags $\vert \Delta \phi \vert
\lesssim 0.4$~rad.  The sub-harmonic usually has a small soft phase lag of the same
magnitude as the fundamental, whereas, the first harmonic has a hard lag, which can be
several times larger than the soft lag displayed by the fundamental.  The coherence of
the fundamental lag is high $\sim$85--95\%, while the coherence of the first and
sub-harmonic are typically closer to 75--80\%.  We identify 50~observations with type
C QPO behavior, only three of which also display high-frequency QPOs (observations 17
\& 18 on 1998 September~20 and observation 161 on 1999 March~13; see Table~1).  

The 50 type C observations can be further subdivided into 45 type C$_1$ and 5 type
C$_2$ classifications.  In the five type C$_2$ observations (Fig.~4 and open circles
in Fig.~6), the fundamental QPO appears to be a superposition of two narrow QPO
features and the harmonics are difficult to discern.  By dividing the observations
into smaller time increments, we found that this blended QPO feature is the result of
a single QPO with a centroid frequency that varies during the observation.  The five
type C$_2$ observations all occurred during the two days (1998 September 20-21)
immediately following the 6.8~Crab flare in the XTE~J1550--564 lightcurve.  The
fundamental type C$_2$ QPOs have frequencies from 6--10~Hz with rms amplitudes of
3--7\% and phase lags of $\sim-0.2$~rad.  When we make the distinction between types
C$_1$ and C$_2$ QPOs, Figure~7a shows that the frequency of type C$_1$ QPOs increases
nearly linearly with the 2--20~keV disk flux.  Two of the three type C observations
that display high-frequency QPOs belong to the C$_2$ subgroup.  

There are two exceptions to the three fundamental types of QPO behavior described
above.  The first anomalous QPO occurred during the 6.8~Crab flare on
19~September 1998 (Sobczak et al.~1999; Remillard et al.~1999).  The PDS for this
observation shows a 13~Hz QPO with an integrated rms amplitude of 1\% (Fig.~5).  The
QPO has a large soft lag of $-0.9\pm0.2$~rad with a coherence of $80\pm20$\%.  The
power-law dominates the spectrum during this flare, with the disk contributing only
$\sim3$\% of the 2--20~keV flux.  This observation is also the first occurrence of the
high-frequency ($\sim183$~Hz) QPO.  The magnitude of the soft lag and the presence of
a high-frequency QPO are consistent with type A behavior, but the QPO frequency and
the low contribution of the disk to the total flux for this observation (Fig.~6) are
similar to type C behavior.  

The second anomalous QPO occurred on 2~March 1999.  This QPO is observed at a much
higher frequency than the others (18~Hz with an indeterminate phase lag and coherence)
with no harmonics.  It is significantly sharper than the other QPOs ($Q \sim 18$) and
has a very small rms amplitude (0.5\%).  The disk component contributed 80\% of the
2--20~keV flux during this observation.  

The three types of QPOs are clearly distinguished by the relative contribution of the
disk component to the total flux (Fig.~6).  However, most of the observations for
which the disk fraction exceeds 0.4 do not exhibit QPOs.  Furthermore, no QPOs have
been detected when the disk fraction exceeeds 0.8.  

The phase lag of the fundamental QPO is plotted vs.~frequency and amplitude in
Figures~8a \& 8b for all the QPO types discussed above.

\section{Discussion}

Hard lags in broadband X-ray lightcurves are often attributed to the Compton
upscattering of photons by hot electrons in an extended corona (Miyamoto et al.~1988;
Hua \& Titarchuk 1996; Kazanas, Hua \& Titarchuk 1997; B\"{o}ttcher \& Liang 1998; see
also Cui~1999 and references therein).  In this scenario, the hard photons undergo
more scatterings in order to reach higher energies and therefore lag behind the soft
photons.  The hard lag is directly related to the photon diffusion time scale through
the corona, which scales logarithmically with photon energy (Hua \& Titarchuk 1996),
in agreement with observations (Cui et al.~1997; Nowak et al.~1999).  The time lags
reported herein for XTE~J1550--564 are typically on the order of 0.001--0.01~s; for a
10~$M_{\odot}$ black hole this corresponds to the light crossing time over 10--100
Schwarzschild radii.  However, the measured lags in other sources can be much larger
(e.g. $\sim1$~s for the 67~mHz (14.9~s) QPO in GRS~1915$+$105 reported by Cui 1999),
which would require an extended Comptonizing region that would be difficult to
maintain physically (Nowak et al.~1999; B\"{o}ttcher \& Liang 1999; Poutanen \& Fabian
1999).  Other models for hard lags include the intrinsic spectral hardening of X-ray
emitting magnetic flares (Poutanen \& Fabian 1999) and waves or blobs of matter moving
inward through an increasingly hotter region where the hard X-rays are emitted
(B\"{o}ttcher \& Liang 1999).  All of these models have been proposed to explain
broadband (continuum) hard lags and do not apply directly to the QPO phenonemon. 
Furthermore, these scenarios cannot explain the observed soft lags, which may be due
to the evolution of the QPO waveform with time (Cui 1999).  In addition, both type B
\& C QPOs exhibit near-perfect coherence, which presents a challenge for QPO models
because there are more mechanisms for destroying coherence than there are for
producing coherence (Vaughan \& Nowak 1997).  

High-frequency QPOs are coincident with all types A and B low-frequency QPOs.  The
low-frequency QPOs are observed at 5--6~Hz in these cases, whereas the high-frequency
QPOs vary in the range 120--285~Hz.  This pattern indicates substantially less
correlation between the high and low-frequency QPOs in black hole sources than that
reported for the corresponding oscillations for accreting neutron stars (van der Klis
et al.~1996; Ford \& van der Klis 1998; Markwardt, Strohmayer, \& Swank 1999). In this
sense, the behavior of XTE~J1550--564 does not support the idea that QPOs in neutron
star and black hole systems are fundamentally related (Psaltis, Belloni, \& van der
Klis 1999).  

We do not include a discussion of the models pertaining to the origin of the QPO
phenomenon.  The reason for this omission is that none of the models proposed to date
contain sufficient detail to fit the observed QPO properties discussed here.  See Cui
(1999), Markwardt, Swank, \& Taam (1999), and Sobczak et al.~(2000a) for a discussion
of QPO models.

\section{Conclusions}

In conclusion, we summarize the observed QPO properties for XTE~J1550--564.  First,
whenever QPOs are observed the power-law component contributes more than 20\% of the
2--20~keV flux (Sobczak et al.~2000a).  During the very high state of the outburst,
when the QPOs have the largest amplitude, the disk component contributes as little as
3\% of 2--20~keV flux (Sobczak et al.~1999,2000b).  However, the QPO frequency
generally increases as the disk flux increases (Fig.~7a).  These results demonstrate
that both the disk and the power-law components are linked to the QPO phenomenon. 
Secondly, there are three fundamental types of QPO behavior which can be grouped
according to the relative contributions of the disk and power-law components to the
total flux (Fig.~6).  The QPOs display both hard and soft lags, as well as near
unity coherence in two of the QPO types.

\acknowledgements G.S. would like to thank J.~Poutanen, J.~Homan, and R.~Wijnands for
helpful discussions.

\newpage
\include{tab1}
\include{tab2}

\newpage

\begin{figure}
\figurenum{1}
\plotone{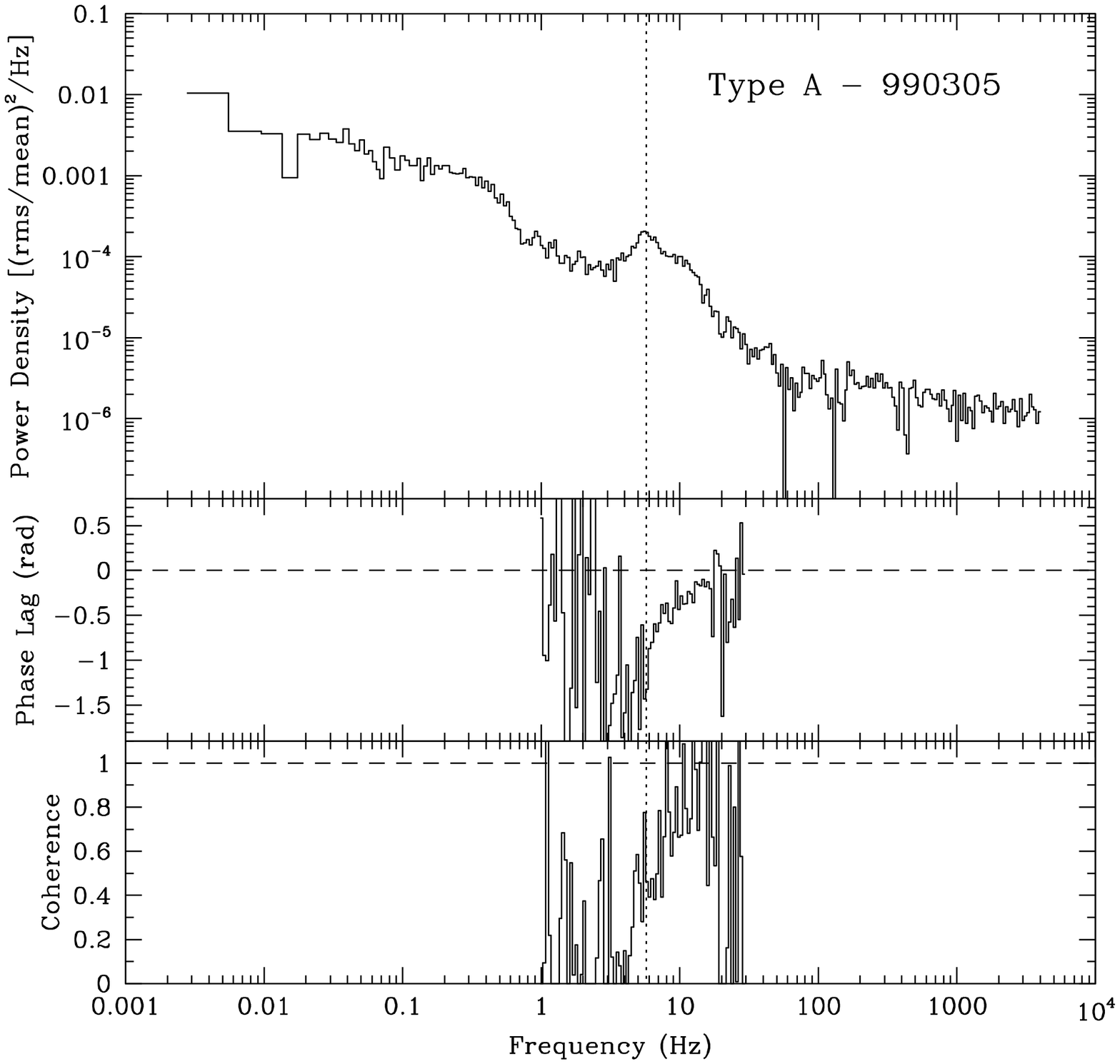}
\caption{Characteristic power spectrum (top), phase lag spectrum (middle), and coherence
function (bottom) for Type~A QPO behavior (see text for details).  The phase lag and coherence
are computed between the 2--13~keV and 13--30~keV bands, with a positive phase lag
representing a hard lag.  For this observation there are $\sim15300$~c/s in the 2--13~keV band and $\sim600$~c/s in the
13--30~keV band.}
\end{figure}

\begin{figure}
\figurenum{2}
\plotone{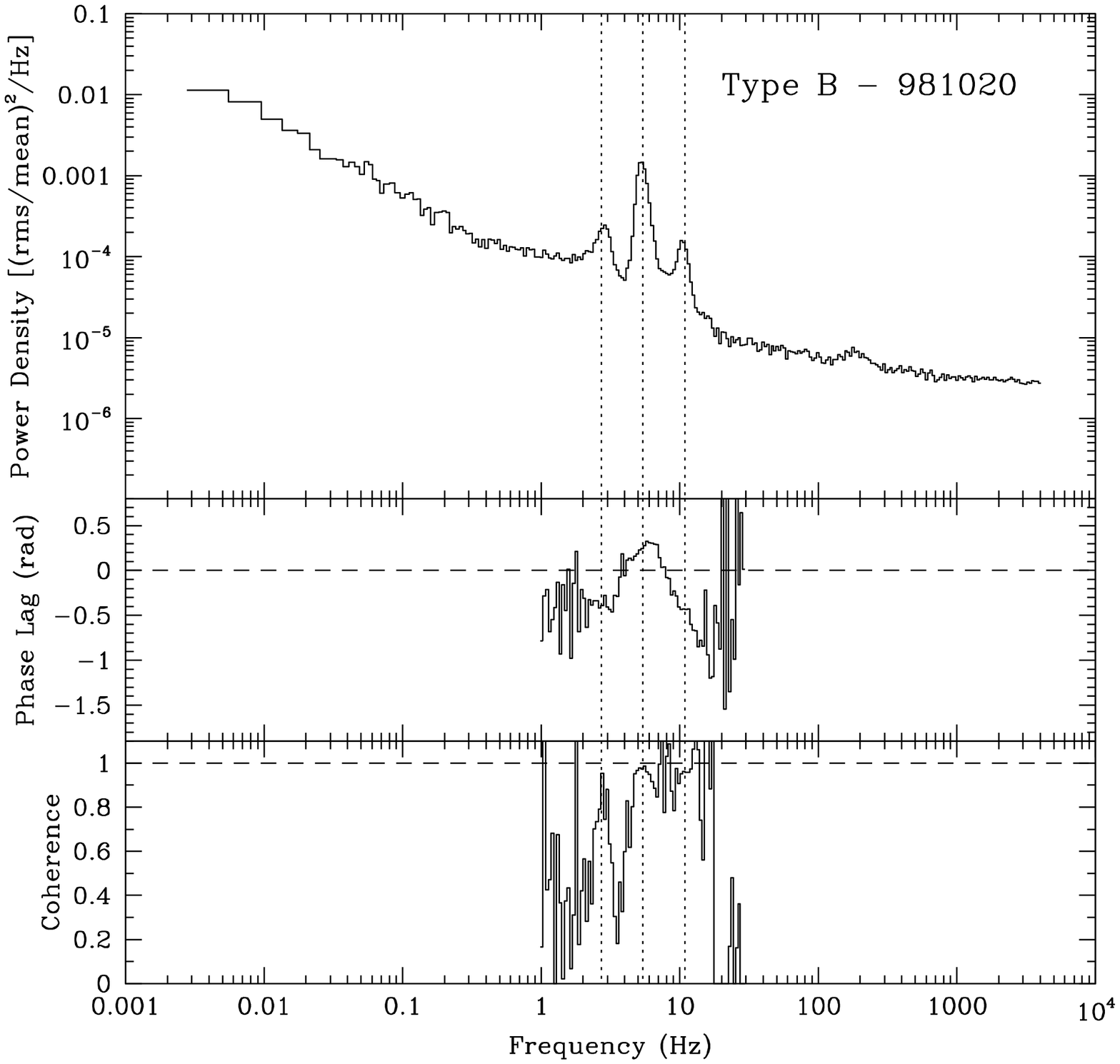}
\caption{Characteristic power spectrum (top), phase lag spectrum (middle), and coherence
function (bottom) for Type~B QPO behavior (see text for details).  
For this observation there are $\sim18200$~c/s in the 2--13~keV band and $\sim1100$~c/s in the
13--30~keV band.}
\end{figure}

\begin{figure}
\figurenum{3}
\plotone{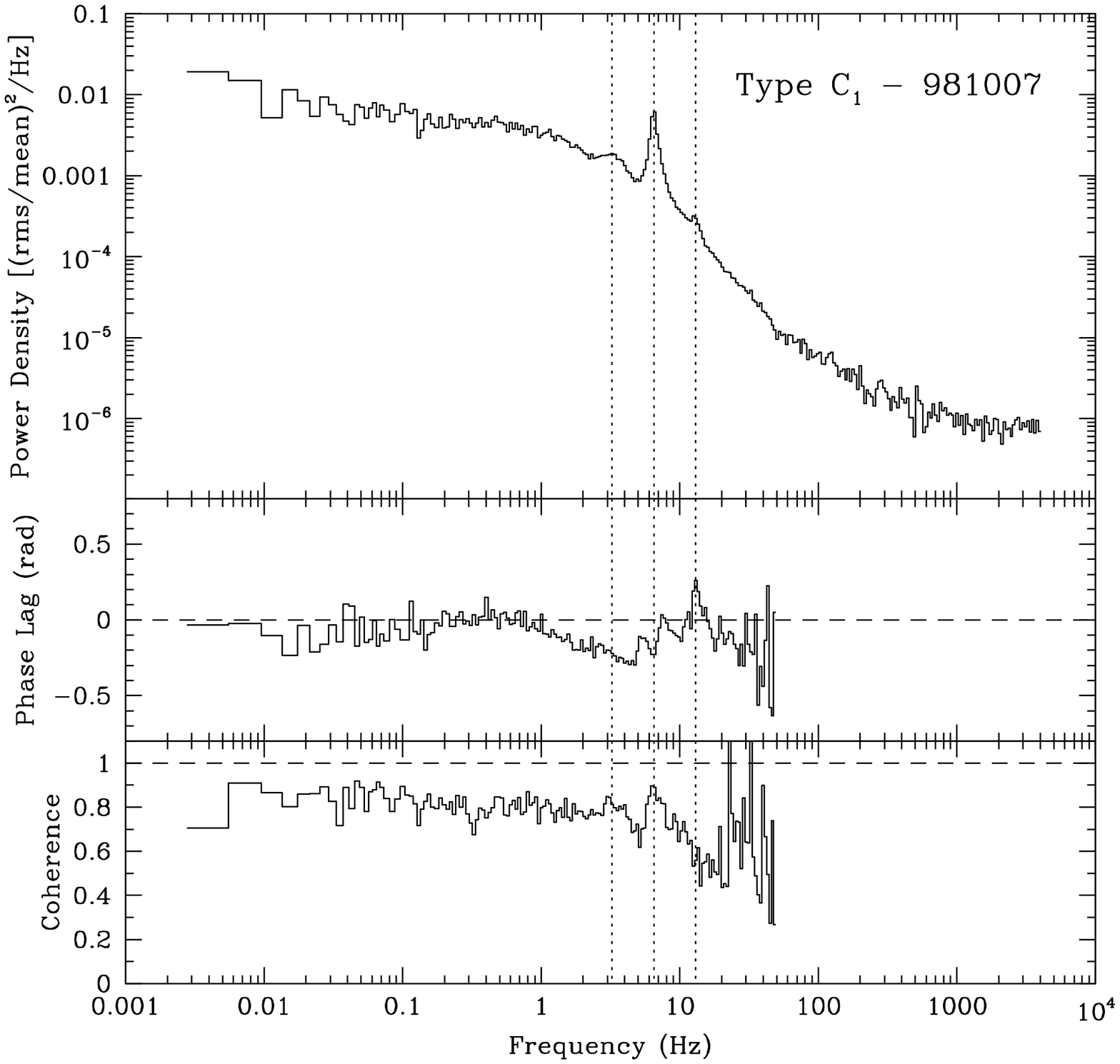}
\caption{Characteristic power spectrum (top), phase lag spectrum (middle), and coherence
function (bottom) for Type~C QPO behavior (see text for details).  
For this observation there are $\sim19000$~c/s in the 2--13~keV band and $\sim1500$~c/s in the
13--30~keV band.}
\end{figure}

\begin{figure}
\figurenum{4}
\plotone{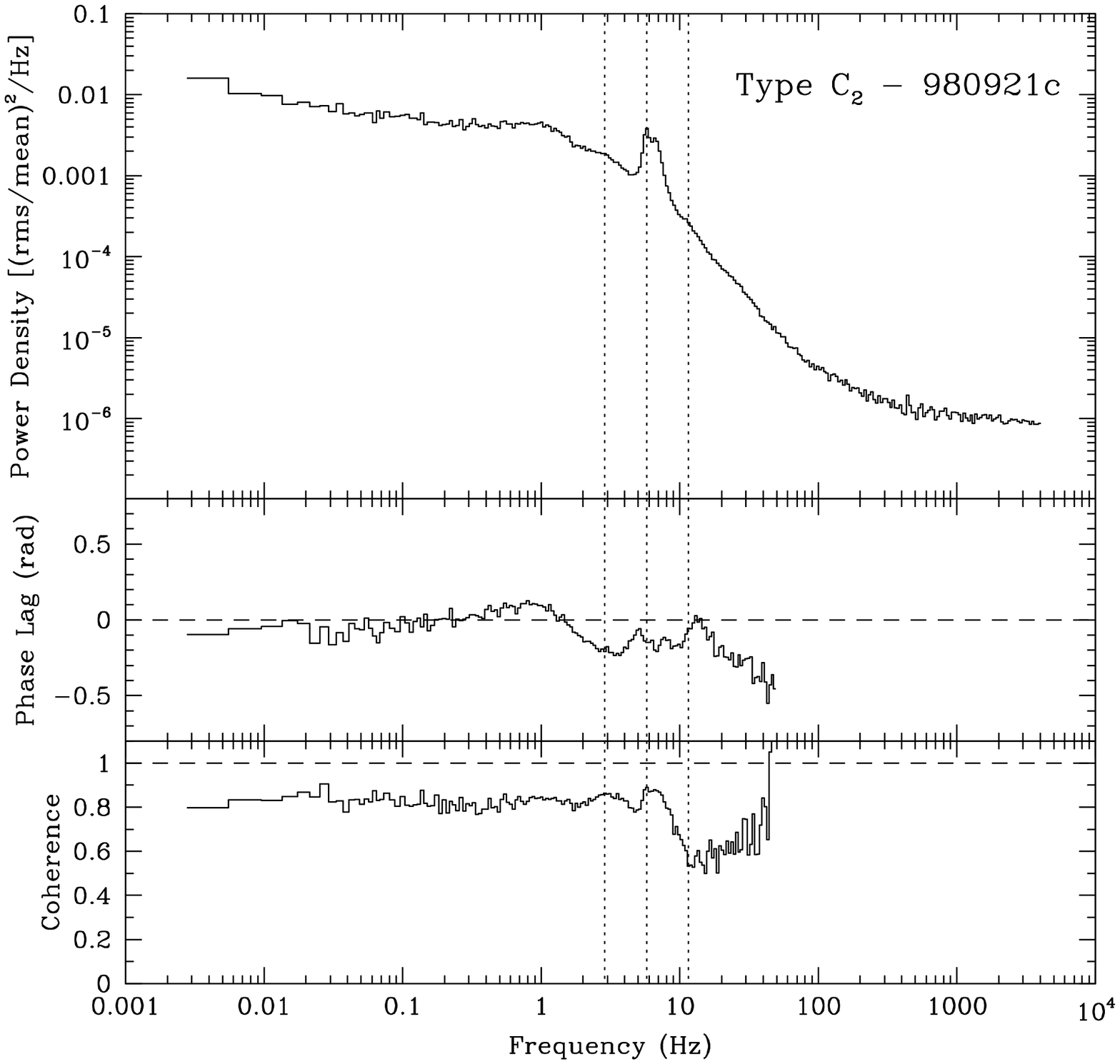}
\caption{Characteristic power spectrum (top), phase lag spectrum (middle), and coherence
function (bottom) for Type C$_2$ QPO behavior (see text for details).  
For this observation there are $\sim27000$~c/s in the 2--13~keV band and $\sim1900$~c/s in the
13--30~keV band.}
\end{figure}

\begin{figure}
\figurenum{5}
\plotone{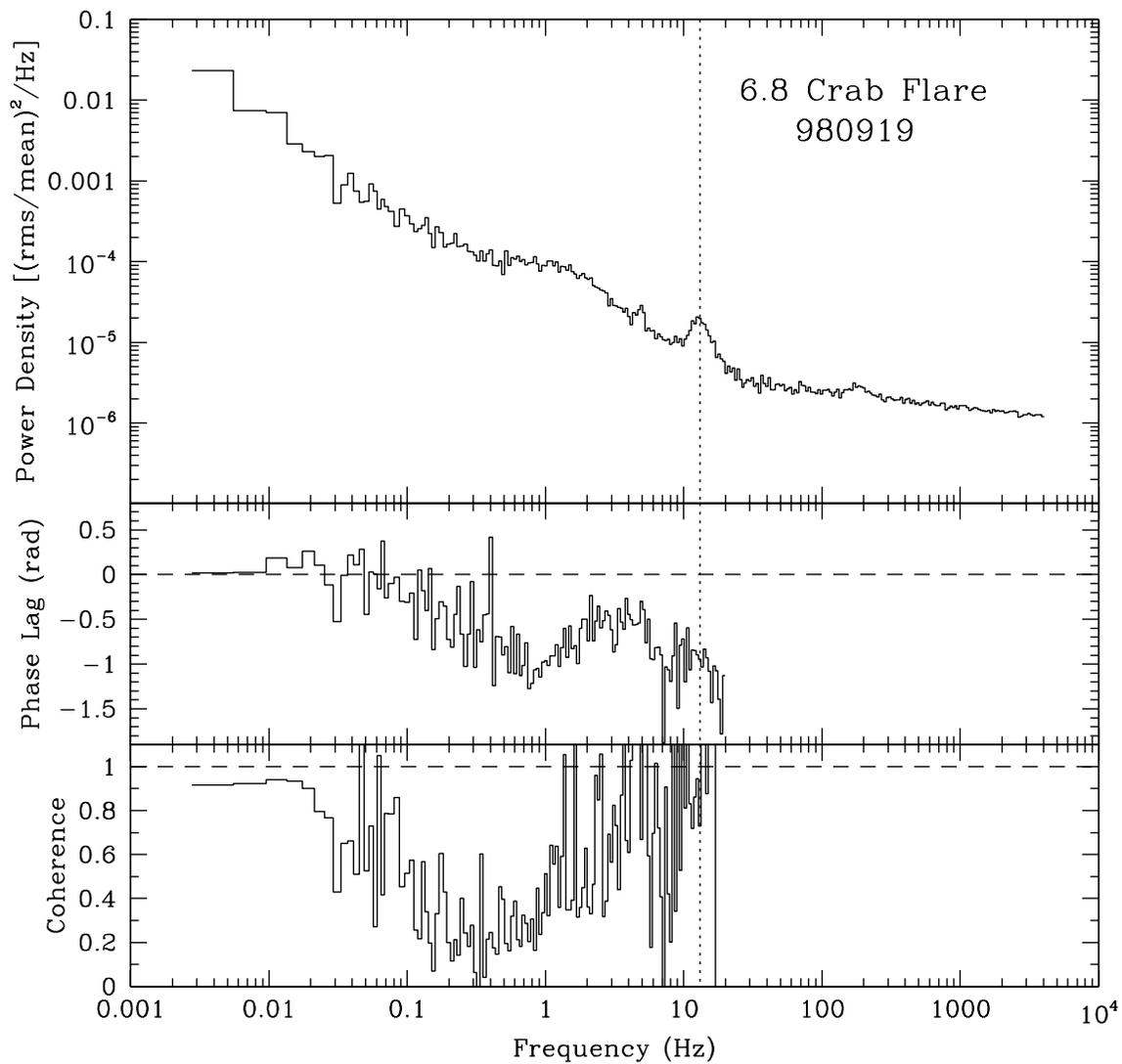}
\caption{Power spectrum (top), phase lag spectrum (middle), and coherence
function (bottom) for 6.8 Crab flare on 19~September 1998, which does not fit the type A, B, C
classification (see text for details).  
For this observation there are $\sim63000$~c/s in the 2--13~keV band and $\sim4100$~c/s in the
13--30~keV band.}
\end{figure}

\begin{figure} 
\figurenum{6} 
\plotone{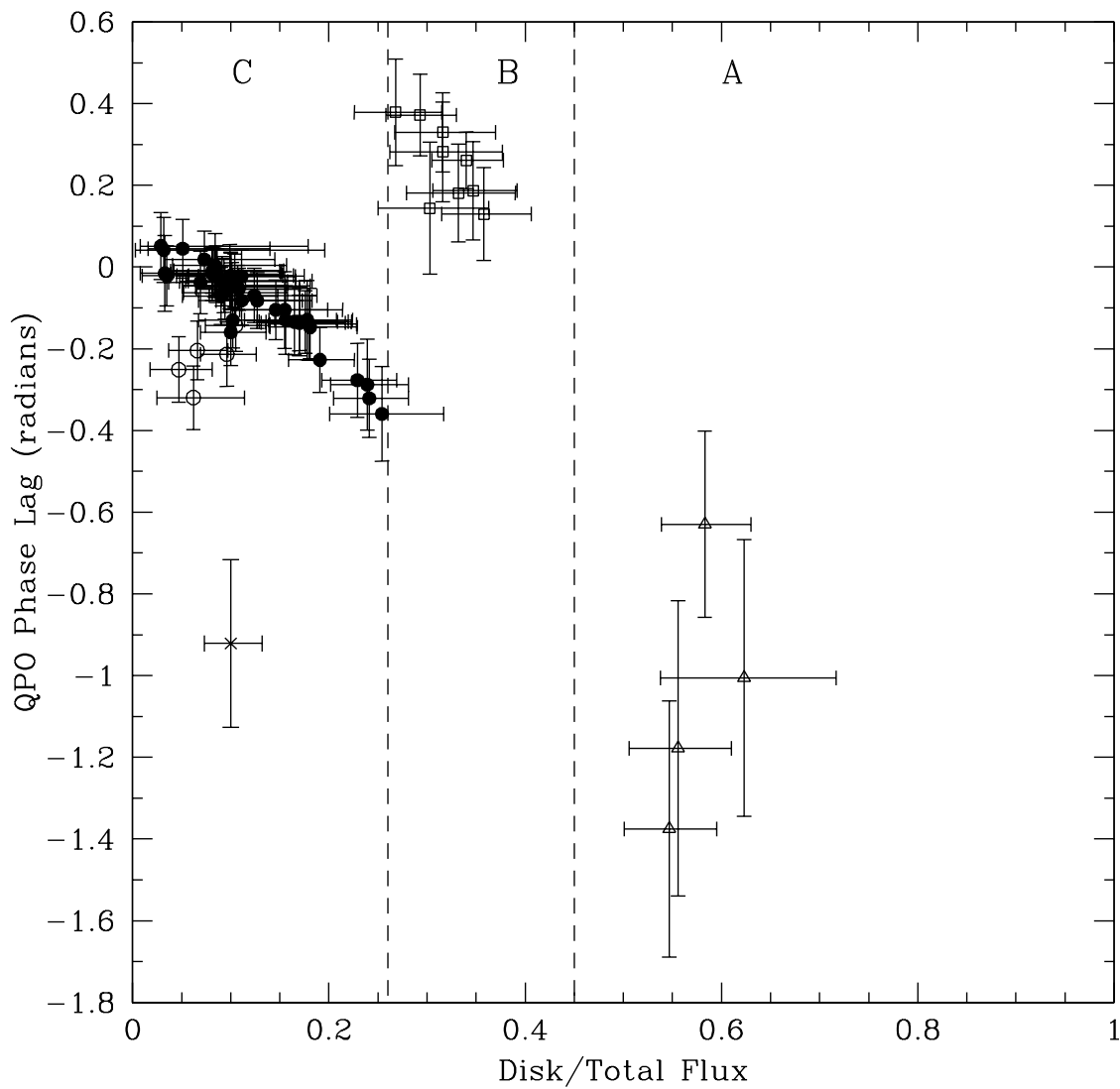} 
\caption{Fundamental QPO phase lag vs. the relative
contribution of the disk component (1 -- power-law component) relative to the total observed
flux from 2--20~keV.  Flux data obtained from Sobczak et al.~(2000b).  The observation
which took place during the 6.8~Crab flare on 19~September 1998 is represented by the
symbol `x'.  Type C$_2$ observations are plotted with open circles.  } 
\end{figure}

\begin{figure}
\figurenum{7}
\plotone{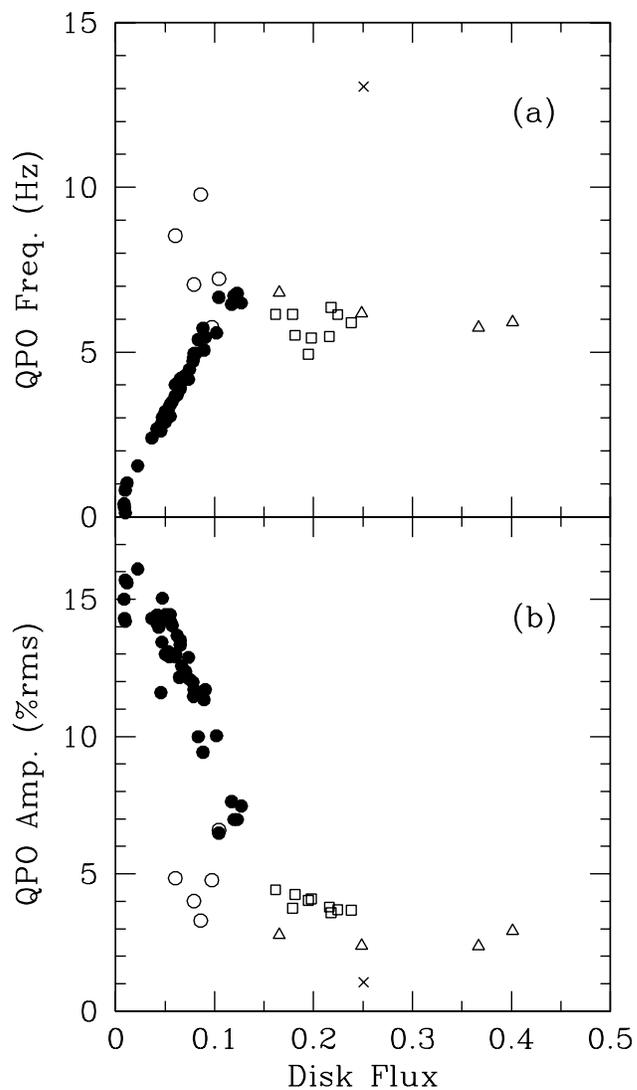}
\caption{(a) QPO frequency and (b) QPO integrated rms amplitude vs. the unabsorbed
2--20~keV disk flux in units of $10^{-7}$~erg s$^{-1}$ cm$^{-2}$.  The fundamental
types of QPO behavior are plotted as follows: Type A -- open triangles, Type B -- open
squares, Type C$_1$ -- filled circles, Type C$_2$ -- open circles, and the 6.8 Crab
flare -- `x'.  The error bars are not shown here, but are comparable to the symbol
size for the QPO frequency and amplitude, and are approximately $\pm0.02$ in the given
units for disk flux.  }
\end{figure}

\begin{figure} 
\figurenum{8} 
\plotone{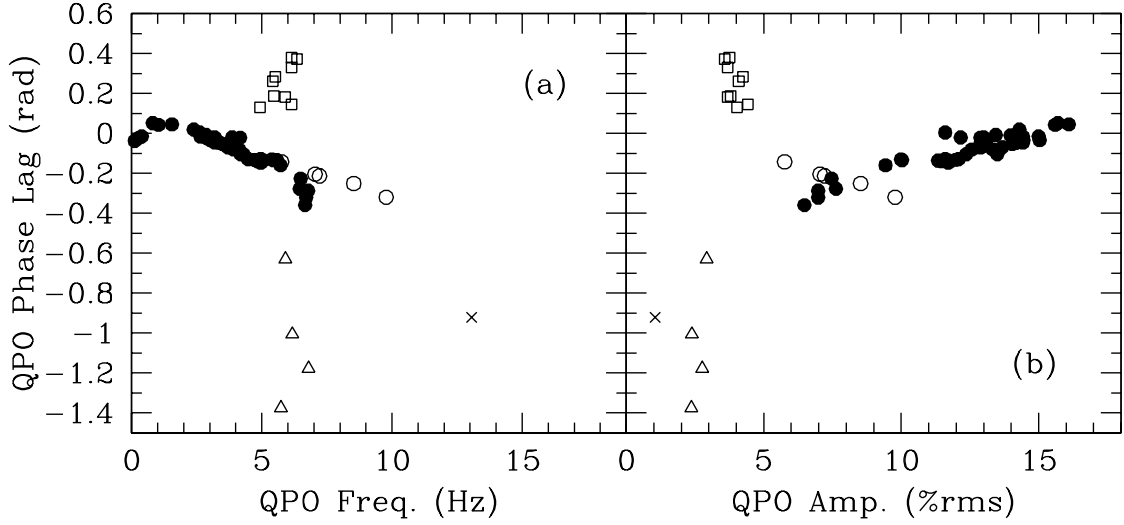} 
\caption{Fundamental QPO phase lag vs. (a) frequency and (b) integrated rms amplitude.
The plotting symbols are the same as in Figure~7.  The error bars are not shown so
that the QPO types are easier to discern.  }
\end{figure}

\end{document}

%% file: tab1.tex


\begin{deluxetable}{lcccccccc}
\scriptsize
\tablewidth{0pt}
\tablenum{1}
\tablecaption{Low-Frequency QPO Parameters for XTE~J1550--564:  Fundamental and Harmonic Frequencies}
\tablehead{
 \colhead{Obs} & \colhead{Date} & \colhead{MJD\tablenotemark{a}} &
 \colhead{Type} & \colhead{QPO Freq.\tablenotemark{b}} & \colhead{QPO Amp.\tablenotemark{c}} & \colhead{Q\tablenotemark{d}} &
 \colhead{Phase Lag\tablenotemark{e}} & \colhead{Coherence\tablenotemark{f}} \cr
 \colhead{\#} & \colhead{(yymmdd)} & \colhead{} &
 \colhead{} & \colhead{(Hz)} & \colhead{(\% rms)} & \colhead{} &
 \colhead{(radians)} & \colhead{} 
}
\startdata
2&  980908&  51064.01&  C&    0.12&  $14.7\pm1.2$&   1.2&  $-0.04\pm0.08$&  $0.99\pm0.03$\nl
\nodata&  \nodata&  \nodata&  \nodata&    0.06&  $ 6.3\pm1.5$&   2.3&  $0.05\pm0.12$&  $0.98\pm0.06$\nl
\nodata&  \nodata&  \nodata&  \nodata&    0.23&  $18.7\pm3.1$&   0.5&  $0.04\pm0.09$&  $0.99\pm0.04$\nl
3&  980909&  51065.07&  C&    0.29&  $16.5\pm1.3$&   9.5&  $-0.02\pm0.07$&  $0.99\pm0.03$\nl
\nodata&  \nodata&  \nodata&  \nodata&    0.16&  $ 5.0\pm1.5$&   4.8&  $-0.03\pm0.12$&  $0.95\pm0.07$\nl
\nodata&  \nodata&  \nodata&  \nodata&    0.59&  $ 7.1\pm0.9$&   7.7&  $0.07\pm0.09$&  $0.97\pm0.05$\nl
4&  980909&  51065.34&  C&    0.39&  $14.7\pm1.5$&   9.9&  $-0.02\pm0.09$&  $0.98\pm0.04$\nl
\nodata&  \nodata&  \nodata&  \nodata&    0.78&  $ 5.7\pm1.2$&  12.1&  $0.08\pm0.12$&  $0.95\pm0.07$\nl
5&  980910&  51066.07&  C&    0.81&  $15.4\pm1.0$&   7.9&  $0.05\pm0.08$&  $0.97\pm0.05$\nl
\nodata&  \nodata&  \nodata&  \nodata&    1.6&  $ 7.6\pm0.5$&   5.5&  $0.17\pm0.11$&  $0.90\pm0.08$\nl
6&  980910&  51066.34&  C&    1.0&  $16.2\pm0.8$&   8.4&  $0.04\pm0.08$&  $0.97\pm0.04$\nl
\nodata&  \nodata&  \nodata&  \nodata&    2.1&  $ 8.1\pm0.4$&   4.7&  $0.17\pm0.11$&  $0.90\pm0.08$\nl
7&  980911&  51067.27&  C&    1.5&  $14.1\pm1.1$&  12.8&  $0.04\pm0.07$&  $0.97\pm0.04$\nl
\nodata&  \nodata&  \nodata&  \nodata&    3.1&  $ 6.6\pm0.3$&   6.8&  $0.18\pm0.11$&  $0.80\pm0.09$\nl
8&  980912&  51068.35&  C&    2.4&  $13.6\pm0.8$&  14.3&  $0.02\pm0.07$&  $0.97\pm0.04$\nl
\nodata&  \nodata&  \nodata&  \nodata&    1.3&  $ 4.4\pm0.7$&   2.5&  $0.02\pm0.14$&  $0.78\pm0.11$\nl
\nodata&  \nodata&  \nodata&  \nodata&    4.7&  $ 5.9\pm0.3$&   8.3&  $0.21\pm0.12$&  $0.78\pm0.10$\nl
9&  980913&  51069.27&  C&    3.3&  $13.1\pm0.6$&  15.3&  $-0.05\pm0.07$&  $0.97\pm0.04$\nl
\nodata&  \nodata&  \nodata&  \nodata&    1.7&  $ 3.5\pm0.5$&   3.4&  $-0.03\pm0.13$&  $0.78\pm0.10$\nl
\nodata&  \nodata&  \nodata&  \nodata&    6.6&  $ 5.1\pm0.2$&   8.9&  $0.27\pm0.13$&  $0.69\pm0.10$\nl
10&  980914&  51070.13&  C&    3.2&  $13.4\pm0.6$&  14.7&  $-0.03\pm0.06$&  $0.97\pm0.03$\nl
\nodata&  \nodata&  \nodata&  \nodata&    1.6&  $ 4.2\pm0.6$&   1.9&  $-0.01\pm0.12$&  $0.80\pm0.09$\nl
\nodata&  \nodata&  \nodata&  \nodata&    6.3&  $ 5.3\pm0.2$&   8.9&  $0.31\pm0.12$&  $0.64\pm0.10$\nl
11&  980914&  51070.27&  C&    3.2&  $13.1\pm0.7$&  18.7&  $-0.02\pm0.07$&  $0.96\pm0.04$\nl
\nodata&  \nodata&  \nodata&  \nodata&    6.3&  $ 5.1\pm0.2$&  11.5&  $0.28\pm0.12$&  $0.68\pm0.10$\nl
12&  980915&  51071.20&  C&    3.7&  $13.0\pm0.4$&  13.3&  $-0.06\pm0.07$&  $0.96\pm0.04$\nl
\nodata&  \nodata&  \nodata&  \nodata&    1.8&  $ 4.4\pm0.6$&   1.7&  $-0.04\pm0.14$&  $0.76\pm0.11$\nl
\nodata&  \nodata&  \nodata&  \nodata&    7.2&  $ 4.8\pm0.2$&  10.0&  $0.28\pm0.13$&  $0.67\pm0.11$\nl
13&  980915&  51072.00&  C&    2.6&  $14.1\pm2.8$&  27.9&  $0.00\pm0.08$&  $0.98\pm0.04$\nl
\nodata&  \nodata&  \nodata&  \nodata&    1.4&  $ 3.4\pm0.7$&   5.9&  $0.00\pm0.17$&  $0.77\pm0.13$\nl
\nodata&  \nodata&  \nodata&  \nodata&    5.1&  $ 5.9\pm0.3$&  12.6&  $0.30\pm0.15$&  $0.75\pm0.12$\nl
\tablebreak
14&  980916&  51072.34&  C&    4.0&  $12.3\pm0.4$&  15.0&  $-0.07\pm0.07$&  $0.96\pm0.04$\nl
\nodata&  \nodata&  \nodata&  \nodata&    2.0&  $ 3.6\pm0.5$&   2.5&  $-0.06\pm0.13$&  $0.80\pm0.10$\nl
\nodata&  \nodata&  \nodata&  \nodata&    7.9&  $ 4.3\pm0.2$&  10.4&  $0.32\pm0.13$&  $0.64\pm0.11$\nl
15&  980918&  51074.14&  C&    5.7&  $ 9.4\pm0.3$&   8.6&  $-0.16\pm0.08$&  $0.92\pm0.05$\nl
\nodata&  \nodata&  \nodata&  \nodata&    2.8&  $ 4.5\pm0.7$&   2.1&  $-0.15\pm0.12$&  $0.86\pm0.09$\nl
\nodata&  \nodata&  \nodata&  \nodata&   11.1&  $ 2.6\pm0.2$&   4.7&  $0.04\pm0.15$&  $0.64\pm0.13$\nl
16\tablenotemark{\dagger,*}&  980919&  51075.99&  ?&   13.1&  $ 1.0\pm0.1$&   2.8&  $-0.92\pm0.21$&  $0.84\pm0.23$\nl
17\tablenotemark{\dagger}&  980920&  51076.80&  C$_2$&    7.2&  $ 6.6\pm0.1$&   5.0&  $-0.21\pm0.08$&  $0.86\pm0.06$\nl
\nodata&  \nodata&  \nodata&  \nodata&    3.3&  $ 4.6\pm0.3$&   1.7&  $-0.27\pm0.10$&  $0.86\pm0.07$\nl
18\tablenotemark{\dagger,1}&  980920&  51076.95&  C$_2$&    8.5&  $ 4.8\pm0.1$&   4.3&  $-0.25\pm0.08$&  $0.88\pm0.06$\nl
\nodata&  \nodata&  \nodata&  \nodata&   16.9&  $ 3.3\pm0.3$&   1.0&  $-0.09\pm0.16$&  $0.71\pm0.18$\nl
19\tablenotemark{1}&  980921&  51077.14&  C$_2$&    9.8&  $ 3.3\pm0.1$&   9.2&  $-0.32\pm0.08$&  $0.90\pm0.05$\nl
\nodata&  \nodata&  \nodata&  \nodata&   20.1&  $ 2.3\pm0.1$&   2.0&  $0.08\pm0.17$&  $0.71\pm0.18$\nl
20\tablenotemark{1}&  980921&  51077.21&  C$_2$&    7.1&  $ 4.0\pm0.1$&   5.4&  $-0.20\pm0.07$&  $0.89\pm0.06$\nl
\nodata&  \nodata&  \nodata&  \nodata&    3.3&  $ 4.8\pm0.2$&   1.4&  $-0.23\pm0.08$&  $0.92\pm0.06$\nl
21&  980921&  51077.87&  C$_2$&    5.8&  $ 4.8\pm0.2$&  11.6&  $-0.14\pm0.06$&  $0.88\pm0.04$\nl
\nodata&  \nodata&  \nodata&  \nodata&    3.0&  $ 4.5\pm0.4$&   1.5&  $-0.19\pm0.08$&  $0.86\pm0.06$\nl
\nodata&  \nodata&  \nodata&  \nodata&   11.4&  $ 2.4\pm0.1$&   2.3&  $-0.08\pm0.11$&  $0.57\pm0.08$\nl
22&  980922&  51078.13&  C&    5.4&  $10.0\pm0.2$&  10.7&  $-0.13\pm0.07$&  $0.92\pm0.04$\nl
\nodata&  \nodata&  \nodata&  \nodata&    2.8&  $ 3.1\pm0.4$&   3.0&  $-0.13\pm0.11$&  $0.85\pm0.08$\nl
\nodata&  \nodata&  \nodata&  \nodata&   10.6&  $ 3.1\pm0.1$&   5.9&  $0.17\pm0.13$&  $0.60\pm0.11$\nl
23\tablenotemark{2}&  980923&  51079.79&  C&    4.2&  $12.9\pm0.4$&   7.3&  $-0.02\pm0.06$&  $0.98\pm0.03$\nl
\nodata&  \nodata&  \nodata&  \nodata&    7.9&  $ 5.6\pm0.3$&   3.4&  $0.08\pm0.10$&  $0.90\pm0.07$\nl
24\tablenotemark{2}&  980924&  51080.08&  C&    3.9&  $12.2\pm0.4$&  13.6&  $-0.02\pm0.06$&  $0.98\pm0.03$\nl
\nodata&  \nodata&  \nodata&  \nodata&    1.9&  $ 4.1\pm0.4$&   2.8&  $-0.05\pm0.10$&  $0.91\pm0.07$\nl
\nodata&  \nodata&  \nodata&  \nodata&    7.6&  $ 4.1\pm0.2$&   9.4&  $0.10\pm0.08$&  $0.90\pm0.06$\nl
25&  980925&  51081.06&  C&    2.9&  $13.4\pm0.4$&   9.8&  $-0.01\pm0.06$&  $0.97\pm0.03$\nl
\nodata&  \nodata&  \nodata&  \nodata&    1.4&  $ 3.9\pm0.5$&   3.8&  $0.02\pm0.12$&  $0.76\pm0.10$\nl
\nodata&  \nodata&  \nodata&  \nodata&    5.7&  $ 6.2\pm0.2$&   5.3&  $0.23\pm0.11$&  $0.73\pm0.09$\nl
26&  980926&  51082.00&  C&    2.7&  $14.0\pm0.4$&  10.4&  $-0.01\pm0.05$&  $0.97\pm0.03$\nl
\nodata&  \nodata&  \nodata&  \nodata&    1.3&  $ 5.2\pm0.7$&   2.4&  $0.04\pm0.11$&  $0.79\pm0.08$\nl
\nodata&  \nodata&  \nodata&  \nodata&    5.4&  $ 5.7\pm0.2$&   7.4&  $0.22\pm0.09$&  $0.75\pm0.08$\nl
27&  980927&  51083.00&  C&    2.7&  $14.4\pm0.4$&   8.9&  $-0.02\pm0.06$&  $0.96\pm0.04$\nl
\nodata&  \nodata&  \nodata&  \nodata&    1.4&  $ 4.6\pm0.8$&   2.9&  $0.02\pm0.12$&  $0.80\pm0.09$\nl
\nodata&  \nodata&  \nodata&  \nodata&    5.2&  $ 6.0\pm0.2$&   7.1&  $0.21\pm0.11$&  $0.75\pm0.09$\nl
28&  980928&  51084.34&  C&    2.7&  $14.1\pm0.5$&  11.0&  $-0.01\pm0.06$&  $0.97\pm0.03$\nl
\nodata&  \nodata&  \nodata&  \nodata&    1.4&  $ 6.6\pm0.9$&   2.2&  $0.02\pm0.12$&  $0.78\pm0.09$\nl
\nodata&  \nodata&  \nodata&  \nodata&    5.3&  $ 5.8\pm0.2$&   6.9&  $0.21\pm0.10$&  $0.75\pm0.08$\nl
29&  980929&  51085.27&  C&    4.1&  $12.6\pm0.3$&  11.1&  $-0.08\pm0.06$&  $0.96\pm0.03$\nl
\nodata&  \nodata&  \nodata&  \nodata&    2.0&  $ 4.7\pm0.3$&   2.7&  $-0.07\pm0.11$&  $0.82\pm0.08$\nl
\nodata&  \nodata&  \nodata&  \nodata&    8.1&  $ 4.0\pm0.2$&   9.0&  $0.23\pm0.11$&  $0.68\pm0.10$\nl
30&  980929&  51085.92&  C&    2.9&  $14.3\pm0.6$&   9.5&  $-0.02\pm0.08$&  $0.96\pm0.04$\nl
\nodata&  \nodata&  \nodata&  \nodata&    1.4&  $ 5.2\pm0.9$&   3.1&  $-0.04\pm0.15$&  $0.77\pm0.12$\nl
\nodata&  \nodata&  \nodata&  \nodata&    5.7&  $ 4.5\pm0.4$&  11.6&  $0.20\pm0.13$&  $0.79\pm0.11$\nl
31&  980929&  51085.99&  C&    3.0&  $14.4\pm0.6$&   6.5&  $-0.03\pm0.07$&  $0.95\pm0.04$\nl
\nodata&  \nodata&  \nodata&  \nodata&    1.5&  $ 5.8\pm0.5$&   2.5&  $0.02\pm0.11$&  $0.80\pm0.09$\nl
\nodata&  \nodata&  \nodata&  \nodata&    6.0&  $ 6.8\pm0.2$&   4.0&  $0.17\pm0.11$&  $0.75\pm0.09$\nl
32&  980930&  51086.89&  C&    3.5&  $14.0\pm0.2$&   8.2&  $-0.05\pm0.05$&  $0.96\pm0.03$\nl
\nodata&  \nodata&  \nodata&  \nodata&    1.7&  $ 6.1\pm0.2$&   2.4&  $-0.00\pm0.09$&  $0.80\pm0.07$\nl
\nodata&  \nodata&  \nodata&  \nodata&    6.9&  $ 5.5\pm0.1$&   5.8&  $0.21\pm0.09$&  $0.74\pm0.08$\nl
33&  981001&  51087.72&  C&    3.4&  $14.2\pm0.3$&   8.9&  $-0.05\pm0.05$&  $0.96\pm0.03$\nl
\nodata&  \nodata&  \nodata&  \nodata&    1.7&  $ 5.8\pm0.3$&   2.8&  $0.01\pm0.10$&  $0.83\pm0.07$\nl
\nodata&  \nodata&  \nodata&  \nodata&    6.8&  $ 5.8\pm0.2$&   5.5&  $0.19\pm0.09$&  $0.76\pm0.08$\nl
34&  981002&  51088.01&  C&    3.2&  $14.4\pm0.4$&   8.6&  $-0.05\pm0.06$&  $0.96\pm0.03$\nl
\nodata&  \nodata&  \nodata&  \nodata&    1.6&  $ 6.4\pm0.4$&   2.3&  $0.00\pm0.11$&  $0.81\pm0.08$\nl
\nodata&  \nodata&  \nodata&  \nodata&    6.3&  $ 6.1\pm0.2$&   5.3&  $0.22\pm0.10$&  $0.76\pm0.08$\nl
35&  981003&  51089.01&  C&    3.0&  $15.0\pm0.5$&   8.2&  $-0.03\pm0.07$&  $0.97\pm0.04$\nl
\nodata&  \nodata&  \nodata&  \nodata&    1.5&  $ 5.9\pm0.6$&   3.2&  $0.00\pm0.13$&  $0.81\pm0.10$\nl
\nodata&  \nodata&  \nodata&  \nodata&    6.0&  $ 6.3\pm0.3$&   5.3&  $0.22\pm0.12$&  $0.79\pm0.10$\nl
36&  981004&  51090.14&  C&    3.9&  $13.3\pm0.4$&  10.2&  $-0.08\pm0.07$&  $0.96\pm0.04$\nl
\nodata&  \nodata&  \nodata&  \nodata&    2.0&  $ 5.7\pm0.3$&   3.0&  $-0.05\pm0.12$&  $0.82\pm0.09$\nl
\nodata&  \nodata&  \nodata&  \nodata&    7.8&  $ 5.3\pm0.2$&   5.7&  $0.23\pm0.12$&  $0.71\pm0.11$\nl
37&  981004&  51090.70&  C&    3.7&  $13.7\pm0.4$&   9.6&  $-0.07\pm0.07$&  $0.96\pm0.04$\nl
\nodata&  \nodata&  \nodata&  \nodata&    1.9&  $ 6.1\pm0.3$&   2.8&  $-0.05\pm0.12$&  $0.83\pm0.09$\nl
\nodata&  \nodata&  \nodata&  \nodata&    7.4&  $ 5.4\pm0.2$&   6.0&  $0.19\pm0.12$&  $0.76\pm0.10$\nl
38&  981005&  51091.74&  C&    5.6&  $10.0\pm0.3$&  11.4&  $-0.13\pm0.08$&  $0.92\pm0.05$\nl
\nodata&  \nodata&  \nodata&  \nodata&    2.8&  $ 4.8\pm0.3$&   2.8&  $-0.13\pm0.13$&  $0.84\pm0.10$\nl
\nodata&  \nodata&  \nodata&  \nodata&   11.0&  $ 3.2\pm0.2$&   6.6&  $0.26\pm0.16$&  $0.67\pm0.17$\nl
39&  981007&  51093.14&  C&    6.5&  $ 7.5\pm0.3$&  11.1&  $-0.23\pm0.08$&  $0.89\pm0.05$\nl
\nodata&  \nodata&  \nodata&  \nodata&    3.3&  $ 4.5\pm0.3$&   2.5&  $-0.23\pm0.12$&  $0.81\pm0.09$\nl
\nodata&  \nodata&  \nodata&  \nodata&   12.7&  $ 2.5\pm0.1$&   4.5&  $0.23\pm0.16$&  $0.55\pm0.14$\nl
40&  981008&  51094.14&  C&    4.3&  $12.4\pm0.4$&  11.1&  $-0.10\pm0.07$&  $0.96\pm0.04$\nl
\nodata&  \nodata&  \nodata&  \nodata&    2.1&  $ 6.2\pm0.3$&   3.0&  $0.00\pm0.12$&  $0.88\pm0.09$\nl
\nodata&  \nodata&  \nodata&  \nodata&    8.6&  $ 4.9\pm0.2$&   6.0&  $0.22\pm0.14$&  $0.78\pm0.15$\nl
41&  981008&  51094.57&  C&    5.1&  $11.3\pm0.3$&  11.2&  $-0.14\pm0.07$&  $0.95\pm0.04$\nl
\nodata&  \nodata&  \nodata&  \nodata&    2.5&  $ 6.0\pm0.2$&   2.6&  $-0.10\pm0.11$&  $0.84\pm0.09$\nl
\nodata&  \nodata&  \nodata&  \nodata&   10.1&  $ 4.1\pm0.1$&   5.9&  $0.23\pm0.13$&  $0.72\pm0.14$\nl
42&  981009&  51095.61&  C&    4.5&  $12.1\pm0.6$&  12.8&  $-0.13\pm0.08$&  $0.96\pm0.05$\nl
\nodata&  \nodata&  \nodata&  \nodata&    2.2&  $ 6.6\pm0.5$&   2.9&  $-0.04\pm0.14$&  $0.84\pm0.11$\nl
\nodata&  \nodata&  \nodata&  \nodata&    8.9&  $ 4.6\pm0.3$&   6.4&  $0.27\pm0.16$&  $0.71\pm0.15$\nl
43&  981010&  51096.57&  C&    5.4&  $11.7\pm0.2$&   4.9&  $-0.14\pm0.08$&  $0.90\pm0.05$\nl
\nodata&  \nodata&  \nodata&  \nodata&    2.7&  $ 7.1\pm0.6$&   2.2&  $-0.11\pm0.11$&  $0.82\pm0.08$\nl
\nodata&  \nodata&  \nodata&  \nodata&   10.0&  $ 7.2\pm0.3$&   1.6&  $0.15\pm0.13$&  $0.60\pm0.11$\nl
44&  981011&  51097.57&  C&    4.7&  $12.0\pm0.5$&  10.3&  $-0.13\pm0.08$&  $0.95\pm0.05$\nl
\nodata&  \nodata&  \nodata&  \nodata&    2.3&  $ 6.3\pm0.4$&   2.8&  $-0.10\pm0.14$&  $0.86\pm0.10$\nl
\nodata&  \nodata&  \nodata&  \nodata&    9.4&  $ 4.5\pm0.3$&   5.9&  $0.28\pm0.16$&  $0.68\pm0.15$\nl
45&  981011&  51097.81&  C&    4.2&  $13.5\pm0.6$&   9.9&  $-0.10\pm0.09$&  $0.95\pm0.06$\nl
\nodata&  \nodata&  \nodata&  \nodata&    2.1&  $ 6.3\pm0.6$&   3.0&  $-0.04\pm0.16$&  $0.80\pm0.12$\nl
\nodata&  \nodata&  \nodata&  \nodata&    8.4&  $ 5.6\pm0.3$&   5.3&  $0.25\pm0.16$&  $0.78\pm0.15$\nl
46&  981012&  51098.28&  C&    5.0&  $11.6\pm0.4$&  10.1&  $-0.13\pm0.10$&  $0.94\pm0.20$\nl
\nodata&  \nodata&  \nodata&  \nodata&    2.5&  $ 6.5\pm0.3$&   3.1&  $-0.11\pm0.14$&  $0.84\pm0.23$\nl
\nodata&  \nodata&  \nodata&  \nodata&    9.9&  $ 4.6\pm0.2$&   5.2&  $0.23\pm0.16$&  $0.79\pm0.26$\nl
47&  981013&  51099.21&  C&    4.8&  $11.5\pm0.5$&  11.1&  $-0.14\pm0.09$&  $0.94\pm0.05$\nl
\nodata&  \nodata&  \nodata&  \nodata&    2.4&  $ 6.2\pm0.4$&   3.2&  $-0.09\pm0.14$&  $0.83\pm0.11$\nl
\nodata&  \nodata&  \nodata&  \nodata&    9.7&  $ 4.6\pm0.2$&   5.5&  $0.33\pm0.16$&  $0.73\pm0.15$\nl
48&  981013&  51099.61&  C&    5.0&  $11.7\pm0.3$&   9.0&  $-0.15\pm0.08$&  $0.94\pm0.05$\nl
\nodata&  \nodata&  \nodata&  \nodata&    2.5&  $ 6.7\pm0.3$&   2.7&  $-0.05\pm0.13$&  $0.82\pm0.09$\nl
\nodata&  \nodata&  \nodata&  \nodata&    9.9&  $ 4.7\pm0.2$&   4.8&  $0.28\pm0.14$&  $0.79\pm0.16$\nl
49&  981014&  51100.29&  C&    6.4&  $ 7.6\pm0.3$&  10.9&  $-0.28\pm0.09$&  $0.89\pm0.06$\nl
\nodata&  \nodata&  \nodata&  \nodata&    3.3&  $ 5.3\pm0.3$&   2.8&  $-0.15\pm0.14$&  $0.82\pm0.11$\nl
\nodata&  \nodata&  \nodata&  \nodata&   12.9&  $ 2.6\pm0.2$&   5.5&  $0.27\pm0.18$&  $0.57\pm0.17$\nl
50&  981015&  51101.61&  C&    6.8&  $ 7.0\pm0.3$&   6.7&  $-0.29\pm0.11$&  $0.84\pm0.08$\nl
\nodata&  \nodata&  \nodata&  \nodata&    3.5&  $ 4.9\pm0.3$&   2.7&  $-0.23\pm0.15$&  $0.77\pm0.12$\nl
\nodata&  \nodata&  \nodata&  \nodata&   13.5&  $ 2.7\pm0.3$&   3.7&  $0.31\pm0.19$&  $0.67\pm0.21$\nl
51&  981015&  51101.94&  C&    6.7&  $ 7.0\pm0.3$&   9.1&  $-0.32\pm0.10$&  $0.88\pm0.07$\nl
\nodata&  \nodata&  \nodata&  \nodata&    3.5&  $ 5.7\pm0.3$&   2.6&  $-0.19\pm0.14$&  $0.78\pm0.11$\nl
\nodata&  \nodata&  \nodata&  \nodata&   13.1&  $ 4.4\pm0.4$&   1.9&  $0.33\pm0.18$&  $0.51\pm0.15$\nl
52\tablenotemark{\dagger}&  981020&  51106.95&  B&    5.5&  $ 3.8\pm0.1$&   8.9&  $0.19\pm0.12$&  $1.01\pm0.09$\nl
\nodata&  \nodata&  \nodata&  \nodata&    2.5&  $ 1.1\pm0.2$&   3.0&  $-0.80\pm0.46$&  $0.27\pm0.53$\nl
\nodata&  \nodata&  \nodata&  \nodata&   10.6&  $ 2.0\pm0.1$&   4.8&  $-0.47\pm0.22$&  $1.11\pm0.31$\nl
53\tablenotemark{\dagger}&  981022&  51108.08&  B&    5.4&  $ 4.1\pm0.0$&   8.6&  $0.26\pm0.07$&  $0.98\pm0.05$\nl
\nodata&  \nodata&  \nodata&  \nodata&    2.8&  $ 1.3\pm0.1$&   5.7&  $-0.33\pm0.20$&  $0.85\pm0.34$\nl
\nodata&  \nodata&  \nodata&  \nodata&   10.5&  $ 2.2\pm0.0$&   5.2&  $-0.43\pm0.11$&  $0.96\pm0.13$\nl
54\tablenotemark{\dagger}&  981023&  51109.74&  B&    4.9&  $ 4.0\pm0.1$&  11.2&  $0.13\pm0.11$&  $0.97\pm0.08$\nl
\nodata&  \nodata&  \nodata&  \nodata&    2.6&  $ 0.5\pm0.2$&   6.6&  $-1.43\pm0.91$&  $0.43\pm1$\nl
\nodata&  \nodata&  \nodata&  \nodata&    9.8&  $ 1.9\pm0.1$&   5.6&  $-0.38\pm0.22$&  $1.14\pm0.33$\nl
59\tablenotemark{\dagger}&  981029&  51115.28&  A&    6.8&  $ 2.8\pm0.2$&   1.8&  $-1.18\pm0.36$&  $0.0\pm1$\nl
151&  990302&  51239.08&  ?&   18.1&  $ 0.5\pm0.1$&  17.6&  $-1.04\pm0.39$&  $0.49\pm0.56$\nl
153\tablenotemark{\dagger}&  990304&  51241.83&  A&    5.9&  $ 2.9\pm0.1$&   3.0&  $-0.63\pm0.23$&  $0.72\pm0.33$\nl
154\tablenotemark{\dagger}&  993005&  51242.51&  A&    5.7&  $ 2.4\pm0.1$&   2.5&  $-1.38\pm0.31$&  $0.62\pm1$\nl
156\tablenotemark{\dagger}&  990308&  51245.35&  B&    6.3&  $ 3.6\pm0.1$&  11.7&  $0.37\pm0.10$&  $0.96\pm0.08$\nl
\nodata&  \nodata&  \nodata&  \nodata&    3.1&  $ 1.8\pm0.1$&   3.6&  $-0.45\pm0.27$&  $1.20\pm1$\nl
\nodata&  \nodata&  \nodata&  \nodata&   12.3&  $ 1.7\pm0.1$&   6.4&  $-0.43\pm0.18$&  $0.98\pm0.22$\nl
158\tablenotemark{\dagger}&  990310&  51247.98&  B&    6.1&  $ 3.7\pm0.1$&   9.9&  $0.33\pm0.10$&  $0.94\pm0.07$\nl
\nodata&  \nodata&  \nodata&  \nodata&    3.1&  $ 2.0\pm0.1$&   4.5&  $-0.51\pm0.21$&  $0.94\pm0.31$\nl
\nodata&  \nodata&  \nodata&  \nodata&   12.0&  $ 1.8\pm0.1$&   5.6&  $-0.41\pm0.17$&  $0.94\pm0.20$\nl
159\tablenotemark{\dagger}&  990311&  51248.09&  B&    5.9&  $ 3.7\pm0.1$&  11.0&  $0.18\pm0.12$&  $0.97\pm0.09$\nl
\nodata&  \nodata&  \nodata&  \nodata&    2.7&  $ 1.4\pm0.2$&   2.4&  $-0.82\pm0.48$&  $0.0\pm1$\nl
\nodata&  \nodata&  \nodata&  \nodata&   11.8&  $ 1.5\pm0.1$&   7.1&  $-0.44\pm0.24$&  $1.16\pm0.35$\nl
160\tablenotemark{\dagger}&  990312&  51249.40&  B&    6.2&  $ 3.8\pm0.1$&   9.4&  $0.38\pm0.13$&  $0.99\pm0.12$\nl
\nodata&  \nodata&  \nodata&  \nodata&    3.1&  $ 2.8\pm0.1$&   5.3&  $-0.27\pm0.23$&  $0.86\pm0.27$\nl
\nodata&  \nodata&  \nodata&  \nodata&   11.9&  $ 1.9\pm0.1$&   5.7&  $-0.48\pm0.24$&  $1.18\pm0.37$\nl
161\tablenotemark{\dagger}&  990313&  51250.69&  C&    6.7&  $ 6.5\pm0.3$&   8.8&  $-0.36\pm0.12$&  $0.82\pm0.10$\nl
\nodata&  \nodata&  \nodata&  \nodata&    3.4&  $ 5.1\pm0.3$&   2.7&  $-0.14\pm0.17$&  $0.71\pm0.16$\nl
\nodata&  \nodata&  \nodata&  \nodata&   13.6&  $ 1.9\pm0.3$&   7.9&  $0.31\pm0.22$&  $0.59\pm0.23$\nl
162\tablenotemark{\dagger}&  990316&  51253.22&  B&    5.5&  $ 4.2\pm0.1$&   9.6&  $0.28\pm0.12$&  $0.98\pm0.10$\nl
\nodata&  \nodata&  \nodata&  \nodata&    2.8&  $ 1.7\pm0.1$&   6.3&  $-0.24\pm0.30$&  $1.33\pm1$\nl
\nodata&  \nodata&  \nodata&  \nodata&   10.6&  $ 2.3\pm0.1$&   5.0&  $-0.39\pm0.22$&  $1.37\pm0.41$\nl
163\tablenotemark{\dagger}&  990317&  51254.09&  B&    6.2&  $ 4.4\pm0.2$&   4.1&  $0.14\pm0.16$&  $0.77\pm0.14$\nl
\nodata&  \nodata&  \nodata&  \nodata&    3.2&  $ 3.3\pm0.2$&   3.1&  $-0.35\pm0.25$&  $0.41\pm0.17$\nl
164\tablenotemark{\dagger}&  990318&  51255.09&  A&    6.2&  $ 2.4\pm0.2$&   1.9&  $-1.01\pm0.34$&  $0.12\pm1$\nl
\nodata&  \nodata&  \nodata&  \nodata&   13.6&  $ 1.8\pm0.3$&   2.1&  $-0.02\pm0.30$&  $0.0\pm1$\nl
\enddata

\tablenotetext{a}{Start of observation, $MJD=JD-2,400,000.5$.}
\tablenotetext{b}{The statistical errors in the QPO centroid frequency are $<1$\% at the 95\% confidence level.}
\tablenotetext{c}{The integrated fractional rms amplitude of the QPO is the square
root of the integrated power in the QPO feature, expressed as a fraction of the mean
count rate.  Errors are given at the 95\% confidence level.}
\tablenotetext{d}{Q = QPO Frequency/FWHM}
\tablenotetext{e}{Phase lag between 2--13~keV and 13--30~keV bands.  A positive value corresponds to a hard lag.  Errors are given at the $1\sigma$ confidence level.}
\tablenotetext{f}{Errors are given at the $1\sigma$ confidence level.}
\tablenotetext{\dagger}{Indicates observations during which high-frequency (100--285~Hz) QPOs are also observed (Remillard
et al.~1999; Homan et al.~2000; Remillard, private communication).}
\tablenotetext{*}{6.8 Crab flare (see Fig.~1 in Sobczak et al.~1999, 2000b).}
\tablenotetext{1}{Soft band: channels 0--30, 2--11.3~keV; Hard band: channels 31--49, 11.3--18.3~keV (see \S2)}
\tablenotetext{2}{Soft band: channels 0--17, 2--6.5~keV; Hard band: channels 18--249, 6.5--30~keV, effectively (see \S2)}

\end{deluxetable}

%% file: tab2.tex
\begin{deluxetable}{lccc}
\scriptsize
\tablewidth{0pt}
\tablenum{2}
\tablecaption{Summary of QPO Types}
\tablehead{
 \colhead{Property} & \colhead{Type A} & \colhead{Type B} & \colhead{Type C}
}
\startdata
Frequency (Hz)&  $\sim6$&  5--6&  0.1--10\nl
Amplitude (\%rms)&  3--4&  $\sim4$&  3--16\nl
Q\tablenotemark{a}&  $\sim$2--4&  $\sim4$&  $\gtrsim10$\nl
Phase Lag (rad.)&  $-0.6$ to $-1.4$&  0 to 0.4&  0.05 to $-0.4$\nl
~~Sub-Harmonic&  \nodata&  soft&  soft\nl
~~1st Harmonic&  soft&  soft&  hard\nl
Coherence&  $<0.5$&  $\sim1$&  $\sim0.9$\nl
HFQPOs&  All&  All&  3/50\nl
\enddata
\tablenotetext{a}{Q = QPO frequency/FWHM}
\end{deluxetable}

%% file: 1550_lags.bbl
\newpage
\begin{references}
\reference{}Bendat, J. \& Piersol, A. 1986, Random Data: Analysis and Measurement
Procedures (New York: Wiley)
\reference{}B\"{o}ttcher, M. \& Liang, E. P. 1998, \apj, 506, 281
\reference{}B\"{o}ttcher, M. \& Liang, E. P. 1999, \apj, 511, L37
\reference{}Cui, W. 1999, \apj, 524, L59
\reference{}Cui, W., Zhang, S. N. \& Chen, W. 2000, \apjl, in press
\reference{}Cui, W., Zhang, S. N., Chen, W., \& Morgan, E. H. 1999, \apj, 512, L43
\reference{}Cui, W., Zhang, S. N., Focke, W., \& Swank, J. H. 1997, \apj, 484, 383
\reference{}Ford, E. C. \& van der Klis, M. 1998, \apj, 506, L39
\reference{}Hua, X.-M. \& Titarchuk, L. 1996, \apj, 496, 280
\reference{}Homan, J., Wijnands, R., van der Klis, M., Belloni, T., van Paradijs, J.,
Klein-Wolt, M., Fender, R., \& M\'{e}ndez, M. 2000, \apj, submitted (astro-ph/0001163)
\reference{}Kazanas, D., Hua, X.-M., \& Titarchuk, L. 1997, \apj, 480, 735
\reference{}Levine, A. M., Bradt, H., Cui, W., Jernigan, J. G., Morgan, E. H.,
Remillard, R., Shirey, R. E., \& Smith, D. A. 1996, \apjl, 469, 33
\reference{}Markwardt, C. B., Swank, J. H., \& Taam, R. E. 1999, \apj, 513, L37
\reference{}Markwardt, C. B., Strohmayer, T. E., \& Swank, J. H. 1999, \apj, 512, L125
\reference{}Miyamoto, S., Kitamoto, S., Mitsuda, K., \& Dotani, T. 1988 \nat, 336, 450
\reference{}Nowak, M. A., Vaughan, B. A., Wilms, J., Dove, J. B., \& Begelman, M. C.
1999, \apj, 510, 874
\reference{}Psaltis, D., Belloni, T., \& van der Klis, M. 1999, \apj, 520, 262
\reference{}Poutanen, J. \& Fabian, A. C. 1999, \mnras, 306, L31
\reference{}Remillard, R. A., McClintock, J. E., Sobczak, G. J., Bailyn, C. D., Orosz,
J. A., Morgan, E. H., \& Levine, A. M. 1999, \apj, 517, L127
\reference{}Smith, D. A. \& RXTE/ASM teams 1998, \iaucirc~7008
\reference{}Sobczak, G. J., McClintock, J. E., Remillard, R. A., Levine, A. M., 
Morgan, E. H., Bailyn, C. D., \& Orosz, J. A. 1999, \apj, 517, L121
\reference{}Sobczak, G. J., McClintock, J. E., Remillard, R. A., Cui, W., Levine, A.
M., Morgan, E. H., Orosz, J. A., \&  Bailyn, C. D. 2000a, \apj, 531, 537
\reference{}Sobczak, G. J., McClintock, J. E., Remillard, R. A., Cui, W., Levine, A.
M., Morgan, E. H., Orosz, J. A., \&  Bailyn, C. D. 2000b, \apj, submitted
\reference{}van der Klis, M., Swank, J. H., Zhang, W., Jahoda, K., Morgan, E. H.,
Lewin, W. H. G., Vaughan, B., \& van Paradijs, J. 1996, \apj, 469, L1
\reference{}Vaughan, B. A. \& Nowak, M. A. 1997, \apj, 474, L43
\reference{}Wijnands, R., Homan, J., \& van der Klis, M. 1999, \apj, 526, 33
\end{references}
